\pdfoutput=1
\PassOptionsToPackage{pdfpagelabels=false}{hyperref}
\expandafter\def\csname ver@fixltx2e.sty\endcsname{}

\documentclass[fleqn,usenatbib,useAMS,usedcolumn]{mnras}

\usepackage{etoolbox}
\makeatletter
\patchcmd\@combinedblfloats{\box\@outputbox}{\unvbox\@outputbox}{}{%
   \errmessage{\noexpand\@combinedblfloats could not be patched}%
}%
\makeatother

\usepackage{newtxtext}
\usepackage{mathptmx}

\usepackage[T1]{fontenc}
\usepackage{ae,aecompl}

\usepackage{graphicx}	% Including figure files
\usepackage{amsmath}	% Advanced maths commands
\usepackage{amssymb}	% Extra maths symbols
\usepackage{latexsym}
\usepackage{url}
\usepackage{siunitx}
\usepackage[british]{babel}

\hypersetup{pdfauthor={{G. Picogna}, {B. Ercolano}, {J.~E. Owen}, {M.~L. Weber}},
            pdftitle={The dispersal of protoplanetary discs I: A new generation of X-ray photoevaporation models},
            pdfkeywords={protoplanetary discs},
            bookmarksnumbered=true}

\graphicspath{{figures/}}

\DeclareSIUnit\au{au}
\DeclareSIUnit\Msun{M$_\odot$}
\DeclareSIUnit\yr{yr}
\DeclareSIUnit\pp{pp}

\title[Photoevaporation models]{The dispersal of protoplanetary discs I: A new generation of X-ray photoevaporation models}

\author[Picogna et al.]{Giovanni Picogna$^{1}$\thanks{E-mail: picogna@usm.lmu.de (GP)}, Barbara Ercolano$^{1,2}$, James~E. Owen$^{3}$, Michael~L. Weber$^{1}$\\
$^{1}$Universit\"ats-Sternwarte, Ludwig-Maximilians-Universit\"at M\"unchen, Scheinerstr. 1, 81679 M\"unchen, Germany\\
$^{2}$Excellence Cluster Origin and Structure of the Universe,
Boltzmannstr.2, 85748 Garching bei M{\"u}nchen, Germany\\
$^{3}$Astrophysics Group, Imperial College London, Blackett Laboratory, Prince Consort Road, London SW7 2AZ, UK}

\date{Accepted 2019 April 2. Received 2019 March 28; in original form 2018 November 23}

\pubyear{2019}

\volume{487}

\begin{document}

\label{firstpage}

\pagerange{\pageref{firstpage}--\pageref{lastpage}}

\maketitle

\begin{abstract}

Photoevaporation of planet-forming discs by high energy radiation from the central star is potentially a crucial mechanism for disc evolution and it may play an important role in the formation and evolution of planetary systems. We present here a new generation of X-ray photoevaporation models for solar-type stars, based on hydrodynamical simulations, which account for stellar irradiation via a significantly improved parameterisation of gas temperatures, based on detailed photoionisation and radiation transfer calculations. This is the first of a series of papers aiming at providing a library of models which cover the observed parameter space in stellar and disc mass, metallicity and stellar X-ray properties. We focus here on solar-type stars (0.7 M$_{\odot}$) with relatively low-mass discs (1\% of the stellar mass) and explore the dependence of the wind mass-loss rates on stellar X-ray luminosity. We model primordial discs and transition discs at various stages of evolution. Our 2D hydrodynamical models are then used to derive simple recipes for the mass-loss rates that are suitable for one-dimensional disc evolution and/or planet formation models typically employed for population synthesis studies. Line profiles from typical wind diagnostics ([OI] \SI{6300}{\angstrom} and [NeII] \SI{12.8}{\micro m}) are also calculated for our models and found to be roughly in agreement with previous studies. Finally, we perform a population study of transition discs by means of one-dimensional viscous evolution models including our new photoevaporation prescription and find that roughly a half of observed transition discs cavities and accretion rates could be reproduced by our models.

\end{abstract}


\section{Introduction}

The modality of disc dispersal is thought to be of fundamental importance to planet formation, yet the responsible mechanism is still largely unconstrained. Photoevaporation from the central star is currently a promising avenue to investigate, but the models developed to date do not yet have enough predictive power for a detailed comparison with the observations. Photoevaporative profiles and rates play a very important role in understanding disc dispersal as well as in the feedback of this process on planet formation and migration \citep[e.g.][]{Ercolano2015,Ercolano2017,Jennings2017,Carrera2017,Jennings2018}.

Previous efforts trying to infer wind mass-loss rates and profiles from observed atomic and ionic emission line intensities and profiles have been inconclusive. These have included observational and theoretical studies of the Ne~II fine structure line at \SI{12.8}{\micro m} \citep{Herczeg2007,Glassgold2007,Pascucci2009,Schisano2010,Ercolano2010}, the [OI] collisionally excited lines at \SI{6300}{\angstrom} and \SI{5577}{\angstrom} \citep{Rigliaco2014,Simon2016,Ercolano2016} as well as a number of other optical collisionally excited metal lines \citep{Natta2014, Canovas2018}. As discussed in \citet{Ercolano2016} collisionally excited lines are unsuitable to infer mass-loss-rates or to constrain the wind driving mechanism due to their strong temperature dependence allowing them only to trace the hot layer of the wind heated by the EUV radiation and not the bulk of the wind. \citet{Ercolano2017} further discuss this issue also in the context of new observational evidence of two distinct components in the emission lines of transition discs, possibly associated with MHD and photo-evaporative winds \citep[see e.g.][]{Simon2016}.

Mid-infrared observations of molecular species (e.g. CO) may provide a viable alternative, and indeed recent observations \citep[e.g.][]{Pontoppidan2011,Brown2013,Klaassen2013,Klaassen2016} have pointed out that these lines may be tracing a slow and partially molecular wind. Current photoevaporative models however lack the ingredients necessary to quantitatively model these types of observations. Specifically, the models of \citet{Owen2010,Owen2011,Owen2012b} and \citet{Alexander2006A,Alexander2006B} perform hydrodynamical calculations of the photoevaporative wind but do not include a chemical network beyond the atomic state. The models of \citet{Gorti2009}, on the other hand, include a thermochemical calculation, but only consider hydrostatic disc atmospheres, thus precluding the calculation of line profiles.
%. Without hydrodynamics, no accurate predictions of wind mass-loss rates and profiles can be made, and absolutely no line profiles can be calculated.
A first step in this direction has been made by \citet{Wang2017} and \citet{Nakatani2018}, where both the hydrodynamics and the chemistry are evolved together, employing a simplified radiative transfer scheme.

We have started a comprehensive effort, in the context of a coordinated program aiming at understanding the nature of transition discs \footnote{DFG Research Unit FOR: 2634/1 "Planet Formation Witnesses and Probes: Transition Discs", \url{www.transitiondiscs.com}}, that finally aims at performing quantitative spectroscopy of disc winds to allow the direct determination of their strength (mass-loss rates and profiles). To achieve these aims we are building new radiation-hydrodynamical calculations of irradiated discs, coupled with photoionisation, chemistry and radiative transfer calculations for a large parameter space, covering stars of different X-ray properties. The comparison of our model grids with existing and upcoming observations will allow us to constrain the mass-loss rates and the launching regions of the wind and thus pin down the underlying driving disc dispersal mechanism.

This paper is the first in a series, where a new, significantly improved, set of hydrodynamical models of X-ray photoevaporation, using an improved temperature scheme are presented. In a following paper, the wind solutions obtained here will be coupled with a chemical network to provide a first assessment of possible molecular diagnostics. As well as providing the basis for our future investigation, this paper specifically aims at addressing a number of questions left open by the widely used models of \citet{Owen2010,Owen2011,Owen2012b,Owen2016}. Namely, the new temperature scheme allows to follow more carefully the evolution of the outer disc during the transition phase addressing the current discrepancy between the observations and the prediction of non-accreting large hole transition discs (\citealt{Owen2012b}, but also see \citealt{Ercolano2018}, W{\"o}lfer et al. 2019, in prep.). The evolution of the gas surface density during the dispersal phase is an important ingredient to planet formation models based on population synthesis \citep[e.g.][]{Mordasini2012}. To that aim, we provide here an accurate scheme that can be easily imported in population synthesis codes.

The paper is organised as follows. In Section~\ref{sec:methods} we describe the numerical methods used and the parameter space studied, focussing on the improvement with respect to previous studies. In Section~\ref{sec:results} we describe the main results, providing analytical fits. Finally, in Section~\ref{sec:conclusions} we draw the main conclusions from this work and outline the future development.

\section{Methods}\label{sec:methods}

Photoevaporative flows originate from gas that is thermally heated from the central star, launched by the pressure gradient and then centrifugally accelerated.
A hydrodynamical calculation of such flows thus requires the calculation of gas temperatures at each hydrodynamical time-step.
Three approaches have been proposed in the literature: (i) use temperature parameterisations, from detailed radiative transfer calculations that solve the heating and cooling equations, for a range of physical and irradiation properties \citep[e.g.][]{Owen2010,Owen2011,Owen2012b,Haworth2016}. (ii) perform a radiative transfer and thermochemical calculation at each hydrodynamical step \citep[e.g.][]{Wang2017, Nakatani2018}. (iii) Alternatively, an analytic model of the self-similar solution of thermal winds topology has been derived for an isothermal disc \citep{Clarke2016}.

In this work, we choose the first approach (see below) as this allows for a much more accurate estimation of the temperatures, which is crucial for our aims. This is because the computational costs of performing a radiative transfer and thermal calculation at each hydrodynamical time-step are so high that extreme simplifications must be adopted in order to make the task viable. %Based on current attempts published in the literature, we find that the simplifications that need to be employed finally lead to unrealistic conclusions.
This point is discussed in more detail throughout the paper, where our results are considered in the context of previous work, including the recent calculations by \citet{Wang2017}.

\subsection{Thermal Calculations}

We have used the gas photoionisation and dust radiative transfer code {\sc mocassin} \citep{MOCASSIN1,MOCASSIN2,MOCASSIN3} to obtain the gas temperature for typical protoplanetary disc conditions, irradiated by the X-ray+EUV spectrum presented by \citet{Ercolano2008,Ercolano2009}.
This synthetic thermal spectrum was generated by the plasma code of \citet{Kashyap2000}, assuming an emission measure distribution based on that derived from RS CVn type binaries by \citet{SanzForcada2002} (which peaks at \SI{d4}{K}) and fits Chandra spectra of T Tauri stars by \citet{Maggio2007}.

We assume solar abundances for the gas \citep{Asplund2005}, depleted for the amount of gas locked in dust grains, according to \citep{Savage1996}.
Possibly, significantly larger amounts of metal depletion are expected to occur and have indeed been reported in recent observations of protoplanetary discs \citep[e.g.][]{Hogerheijde2011,Favre2013,Ansdell2016,Kama2016,Du2017,Miotello2017}.
The effect of metal depletion on photoevaporation rates can be severe \citep{ErcolanoClarke2010} and can have dramatic effects on the surface density evolution of discs \citep{Ercolano2018}.
A detailed investigation of this effect on the wind rates and profiles by means of hydrodynamical simulations similar to those presented here is currently underway and will be presented in a forthcoming paper (W\"olfer et al., in preparation).

We considered \num{10} different slabs of \SI[mode=text]{2.5d21}{pp.cm^{-2}} each, up to a column density of \SI[mode=text]{2.5d22}{pp.cm^{-2}}, and obtained for each a gas temperature prescription. The column is resolved in 100 grid points and we checked that larger resolution does not change the result.
Beyond the maximum column density, we assume thermal coupling of the gas and dust, and the dust temperatures from the models of \citet{DAlessio2001} are mapped to our models.

The models consist of 3D slabs with reflecting boundary conditions on the xy and zy planes and outflowing boundary conditions on the xz planes. The energy packets are injected plane parallel, but they are allowed to be absorbed, re-emitted and scattered in 3D. The main heating channel is X-ray photoionisation, although photoelectric heating from dust grains is also included, but is mostly negligible. Cooling channels include continuum process (2-photon continuum, free-bound recombination, free-free), however the dominant cooling mechanisms in these models are collisionally excited (forbidden) lines and recombination lines, which are generally optically thin and escape without further interactions, except in the case of resonant emission lines, which could be absorbed by dust and are treated using and escape probability approach (see Ercolano et al. 2005). HeI and HeII recombination lines (the latter not relevant in these models) emitted at energies $>$ 13.6eV are followed as they could go on ionising H or HeI respectively.

Based on our model results we propose a new parameterisation of the gas temperature as a function of ionisation parameter $\xi = L_X/nr^2$ \citep{Tarter1969} and gas column density.

The following relation produces a good fit to the model data:
\begin{equation}
	\log_{10}(T(\xi)) = d + \frac{1.5-d}{\left[1.0 + (\log_{10}{(\xi)}/c)^b\right]^m}\,
\end{equation}
where the coefficients, as a function of the hydrogen column densities $N_H$, are given in Table~\ref{tab:fitTemp}.

\begin{table}
\caption{Coefficients for the Temperature fit}
\label{tab:fitTemp}
\begin{tabular}{lcccc}
\hline
\hline
$N_H$ & b & c & d & m\\
\SI{1d20}{pp/cm^2} & & & &\\
\hline
$0 - 25$ & $-26.5735$ & $-6.5924$ & $4.0359$ & $0.2086$ \\
$25 - 50$ & $-53.6260$ & $-5.8336$ & $3.9421$ & $0.1220$ \\
$50 - 75$ & $-65.0650$ & $-5.7396$ & $3.9160$ & $0.1027$ \\
$75 - 100$ & $-33.4958$ & $-5.4189$ & $3.9088$ & $0.2039$ \\
$100 - 125$ & $-32.0490$ & $-5.2748$ & $3.8974$ & $0.1972$ \\
$125 - 150$ & $-35.0241$ & $-5.1722$ & $3.8840$ & $0.1699$ \\
$150 - 175$ & $-40.8293$ & $-5.1922$ & $3.8705$ & $0.1422$ \\
$175 - 200$ & $-36.0245$ & $-5.2231$ & $3.8720$ & $0.1527$ \\
$200 - 225$ & $-54.3903$ & $-5.0610$ & $3.8584$ & $0.1053$ \\
$225 - 250$ & $-36.0854$ & $-4.9851$ & $3.8635$ & $0.1579$ \\
\hline
\end{tabular}
\end{table}

\begin{figure}
\centering
\includegraphics[width=.45\textwidth]{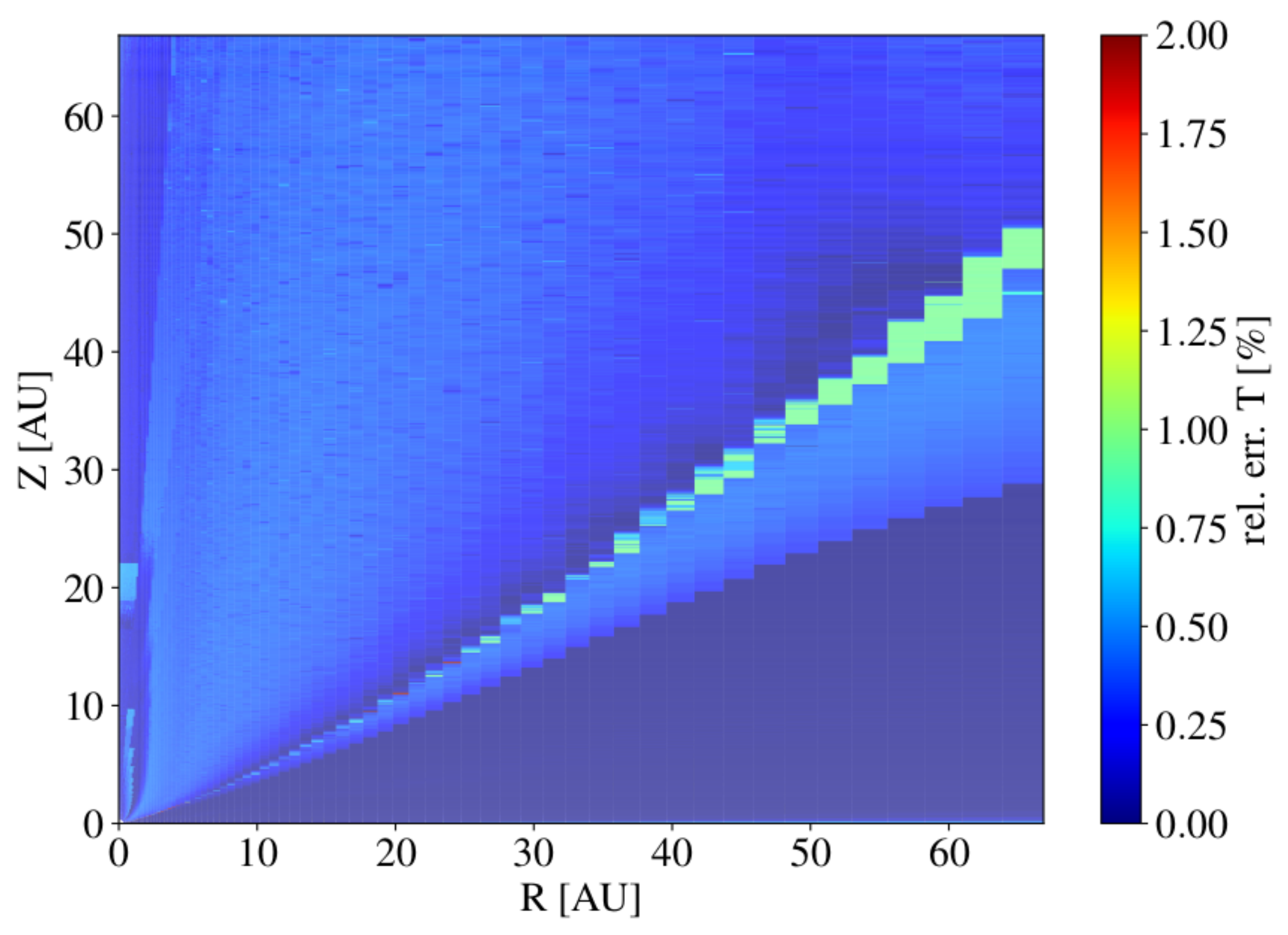}
\caption{Relative error of the temperature determined in PLUTO from the fits, with respect to the one post-processed with MOCASSIN after a quasi-equilibrium state is reached.\label{fig:tempcomp}}
\end{figure}

\begin{figure}
\centering
\includegraphics[width=.45\textwidth]{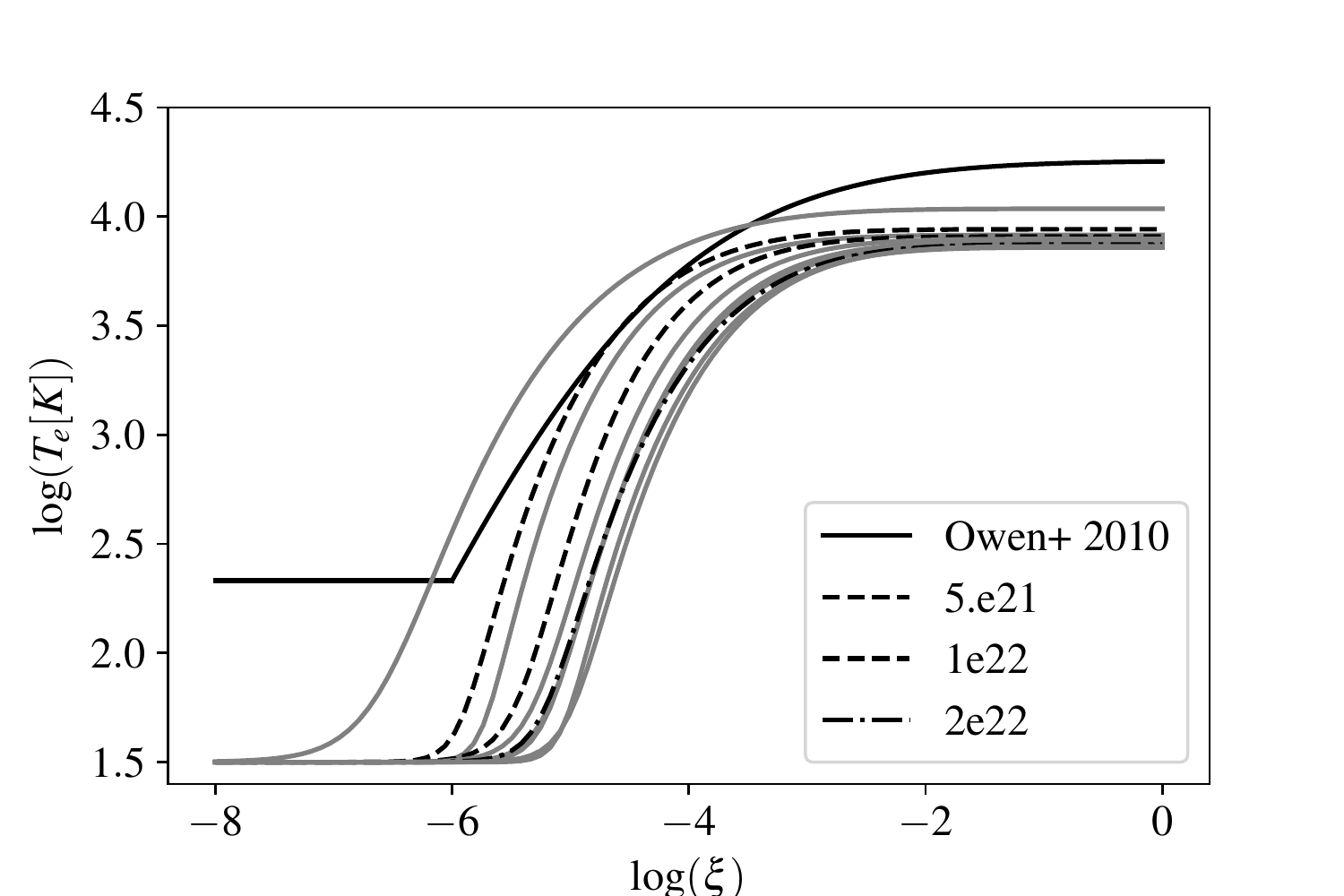}
\caption{Temperature as a function of the ionization parameter for $3$ selected column densities, plotted with grey lines. A comparison with the parametrization by \citet{Owen2010} is given by a black solid line.\label{fig:tempxi}}
\end{figure}

Figure~\ref{fig:tempxi} shows a comparison of our parameterisation scheme to the one used by \citet{Owen2010,Owen2011,Owen2012b}, which was independent of the gas column density, and is likely a column density biased average.
Our new radiative transfer and thermal calculations extend to much lower values of $\xi$ (down to $\log{\xi} = -8$, instead of $\log{\xi} = -6$), allowing us to better model the outer regions of the disc, which are particularly relevant to follow the evolution of transition discs.
The higher maximum temperature reached by the previous prescription (as seen in the upper right part of Fig.~\ref{fig:tempxi}) is due to an integration over a finer grid, which allowed them to solve the low-density region heated by the EUV radiation. We chose not to over-resolve this region since it does not contribute to the total mass-loss rate.
Including the column density as a parameter, allows us to better reproduce the temperatures at different locations of the disc. While the single-slab parameterisation of \citet{Owen2010,Owen2011,Owen2012b} can produce errors of order \SI{30}{\percent} for column densities below \SI[mode=text]{d22}{pp.cm^{-2}}, we reduce this value to less than \SI{1}{\percent} as shown in Figure~\ref{fig:tempcomp}
The consequences of the new temperature prescription on the photoevaporation rates and profiles are discussed in more details in Section~\ref{sec:results}.

Finally, \citet{Haworth2016} presented a parameterisation scheme of temperature as a function of ionisation parameter to model the very late evolution of transition disc and the onset of the thermal sweeping instability \citep{Owen2013}.
We draw attention to the fact, also discussed in \citet{Haworth2016}, that the temperature ionisation prescription presented in their work (see their Figure~1) is only valid under the assumptions made there. \citet{Haworth2016} used gas that was optically thin to low energy X-ray photons to compute the temperatures and assumed that the profile is independent of column density. These assumptions are not justified in our disc models.
%Based on their parameterisation they concluded that thermal sweeping is unimportant to the late evolution of discs.
%In this work we have used our parameterisation to checked that their conclusions remain unchanged when using the new parameterisation presented in this work (T.~Haworth, private communication).

\subsection{Hydrodynamics}\label{sec:hydro}

We used the open source, modern hydro-code \textsc{PLUTO} \citep{PLUTO}, adopting a spherical coordinate system centred onto the star.
This choice of the reference frame allows us to determine, without increasing the computational load, the column density for each cell in the domain.
We then modified the temperature calculation in the code in order to accurately model the stellar irradiation in the upper layers of the disc, using the parameterisation presented in the previous section, without impacting the code performance.

The initial density and temperature distributions for the hydrodynamical simulation of the primordial disc were taken from the hydrostatic equilibrium models of \citet{Ercolano2008,Ercolano2009}.
We note that the choice of initial distribution does not affect the final result, but has an impact on the computational time required to achieve a steady state solution of the wind.
In order to avoid numerical issues due to the very low density in the region near the pole and for large radii, the grid has been defined with an inverse logarithmic scaling in the polar direction, and a logarithmic scaling in the radial one, taking advantage of the polar coordinates.

Particular attention has been made to correctly model the outer boundary of the computational domain that may cause spurious oscillations (observed also in \citet{Wang2017}), and that may affect the final result.
In order to damp spurious numerical reflection, we adopted an outer boundary inside the computational domain. Practically, we set the right-hand side of the conservative equations in the region outside \SI{980}{\au} to zero.  These cells are thus not evolved in time.
A more detailed analysis of this issue is included in Appendix~\ref{app:outerradius}.

The transition discs were derived from the primordial disc distribution by adding an exponential decay of the density close to the gap edge, from the hydrostatic models of \citet{Ercolano2008,Ercolano2009}.
The disc is then readjusting its shape close to the gap as the system reaches a stable state where the gas inflow is balanced by the photoevaporative wind.

We checked the consistency of our results by increasing the grid resolution for our fiducial model (PR).
The list of the systems modelled is reported in Table~\ref{tab:massLoss}.

The parameter space analysis performed is summarised in Tab.~\ref{tab:parSpace}.

\begin{table}
\caption{Parameter space \label{tab:parSpace}}
\centering
\begin{tabular}{l c}
\hline
\hline
\bf{variable} & \bf{value}\\
\hline
\textit{disc extent} & \\
\hline
radial [au] & $0.33-1000$ \\
polar [rad] & $0.005 - \pi/2$ \\
\hline
\textit{grid resolution} & \\
\hline
radial & $412$ \\
polar & $160$ \\
\hline
\textit{physical properties} & \\
\hline
$\log_{10}(L_\mathrm{X})$ [erg/s] & $28.3,29.3,29.8,30.3,31.3$ \\
$R_\mathrm{hole}$ [au] & $6.2,9.9,17.9,28.2$\\
$\mu$ & $1.37125$\\
\hline
\end{tabular}
\end{table}

\subsection{Mass-loss rates calculations}\label{sec:massloss}

We derive the mass-loss rates as a function of the cylindrical radius in the disc, adopting a similar approach as in \citet{Owen2010}.
First, we remapped the grid onto a Cartesian grid \num{2000 x 2000} in extent.
Then, we defined an outer radius and followed the gas streamlines from it to the base of the flow, obtaining the radius at which the flow was launched.
The outer radius was chosen to be \SI{200}{au} in order to avoid the outermost regions, affected by numerical oscillations caused by the reflection of sound waves onto the outer boundary, and maximise the number of orbits performed at the given location within the simulation (see Appendix~\ref{app:outerradius} for more details).

The flow base is defined at the location where a local maximum is observed in the temperature profile for each cylindrical radius.
We checked that this definition was consistent also with the Bernoulli parameter \citep[e.g.][]{Wang2017}.

%{\color{red} JAMES: please review the following paragraph, do you agree with this? do you want to add anything?\\}
%\citet{Wang2017} claim that some of the differences they find with \citet{Owen2010} calculation of the mass-loss rates and wind profiles are to be attributed to their applying a Bernoulli mask to their grids, prior to the calculation of the mass-loss rates.
%We note here that \citet{Owen2010} performed a-posteriori checks of the Jacobi potential, the analogue of the Bernoulli potential in a rotating frame, along the streamlines of their flow, finding that this is conserved to within \SIrange[range-units = single]{3}{4}{\percent}.
%Thus we exclude that the differences with \citet{Owen2010} highlighted by \citet{Wang2017} can be attributed to the Bernoulli mask they apply.
%We have applied a similar procedure here, and we find that in our grids the Jacobi potential is conserved as well to within \SI{4}{\percent}, indicating that the flow is close to steady-state.

\section{Results}\label{sec:results}

In this section, the results from our calculations are described in the context of previous analysis and the consequences of our improved temperature scheme on mass-loss rates and wind profiles are investigated.
We provide, where possible, simple analytical formulae for implementation into disc evolution and planet formation models, which also include dependences on X-ray luminosities, as well as on the inner hole size in the case of transition discs.
Transition disc demographics deriving from our new models are also calculated and compared with previous calculations and with current observations.
Classical wind tracers, including the Ne~II fine structure line at \SI{12.8}{\micro m} and the collisionally excited neutral oxygen line at \SI{6300}{\angstrom} are calculated for our fiducial model and compared to previous calculations by \citet{Ercolano2016} and to  observations.
%Furthermore, we perform the analysis of several collisionally excited neutral Carbon lines at \SIlist[list-units = single]{492; 809; 304476.9; 305286.9; 343661.4}{\GHz} \citep{Abrahamsson2007}.
The raw model grids for each individual case are available from the authors upon request.

\subsection{Mass-loss rates and profiles}

\begin{figure*}
\centering
\includegraphics[width=.95\textwidth]{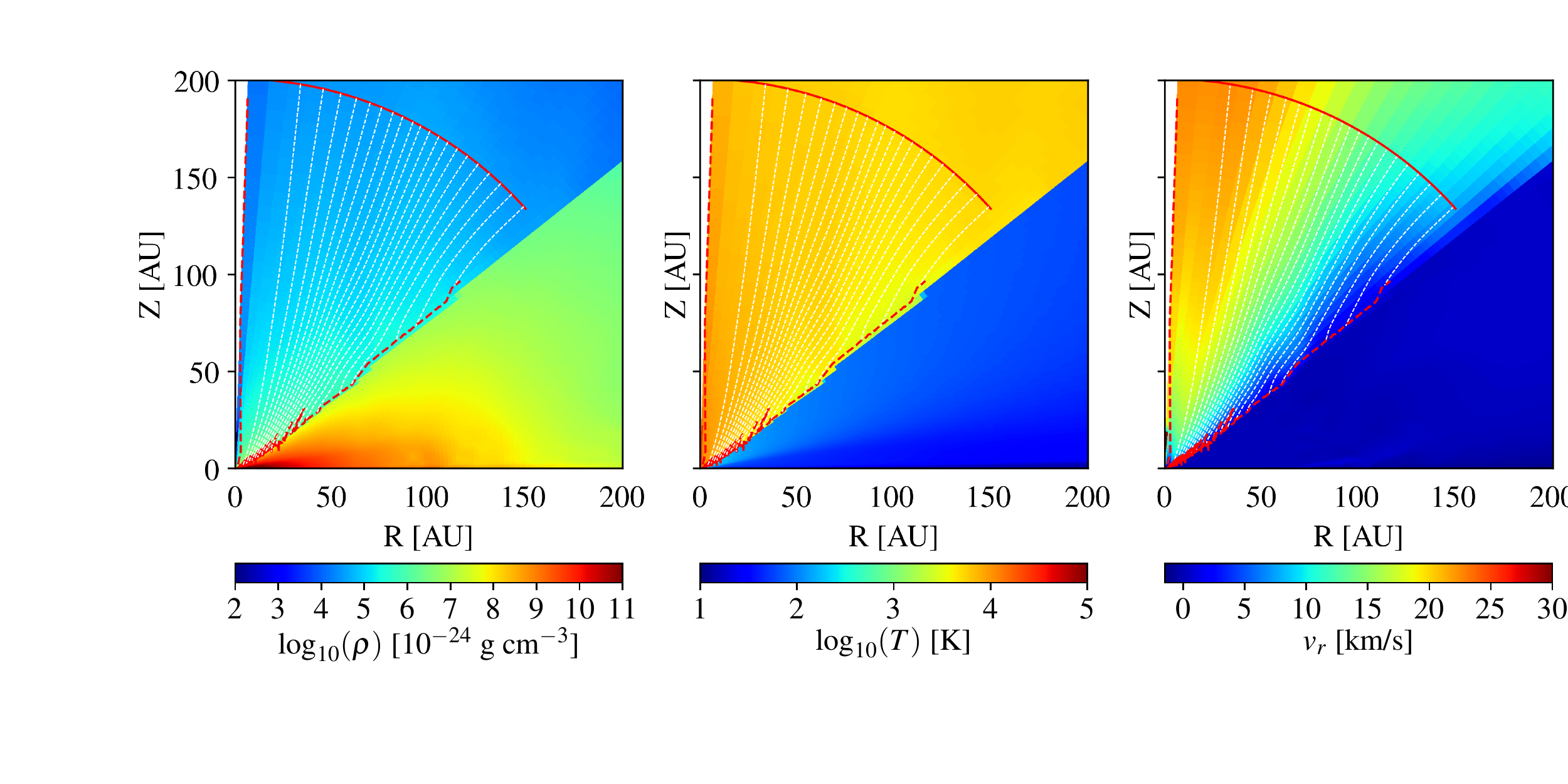}
\caption{Meridional plots for the fiducial model (PR). Left panel: mass density; middle panel: temperature; right panel: radial velocity.
The streamlines are plotted with solid white lines at 5 per cent intervals of the integrated mass-loss rate, and the sonic surface is in plotted with a dashed red line. The surface where the integrated mass-loss rates are computed is highlighted with a solid red line. \label{fig:SigmaVelTemp}}
\end{figure*}

Fig.~\ref{fig:SigmaVelTemp} shows the density, temperature and radial velocity plots at the end of the simulation for the primordial disc in our reference model (PR).
The streamlines of the wind flow are overlaid and plotted every $5 \%$ intervals of the integrated mass-loss rate, from the outer radius, placed at \SI{200}{\au} (well inside the computational domain), down to the base of the flow.
We can see that they originate mostly inside \SI{50}{\au} and they are radially spreading at larger radii as also seen by \citet{Owen2010,Wang2017} as well as in pure EUV models \citep{Font2004,Alexander2006A,Alexander2008}.
The red dashed line marks the sonic surface, the smoothness of which indicates that we are in a stable state condition.

\begin{figure}
\centering
\includegraphics[width=.45\textwidth]{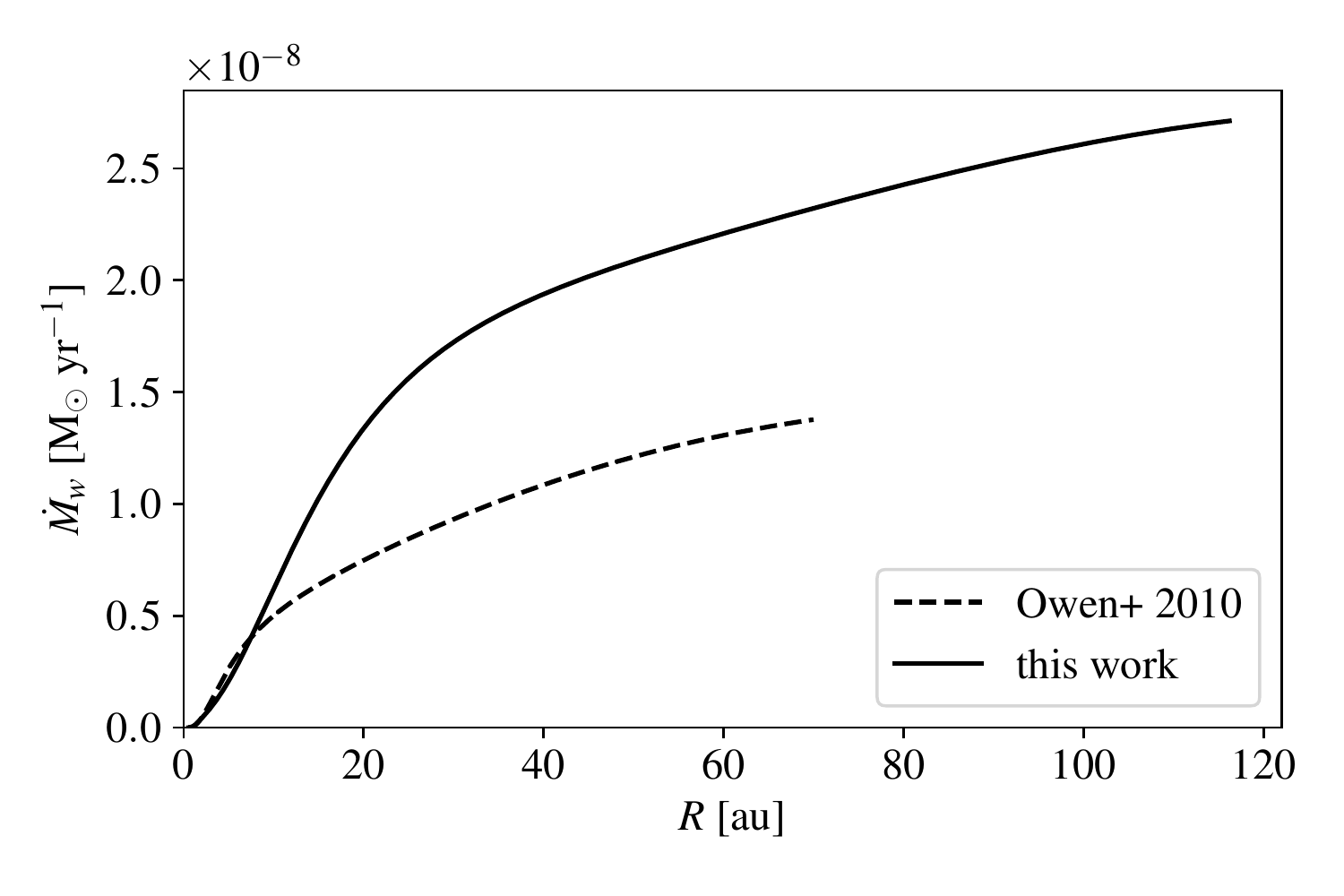}
\caption{Cumulative mass-loss rate as a function of cylindrical radius compared with the prescription of \citet{Owen2010} with the dashed line. \label{fig:CumMdot}}
\end{figure}

The differences with the parameterisation by \citet{Owen2010} are subtle but important and are highlighted in Fig.~\ref{fig:CumMdot}.
The new cumulative mass-loss rate is consistently higher between $5$ and \SI{50}{\au}, boosting the total mass-loss rate which is reported in Tab.~\ref{tab:massLoss} and is $2$ times bigger than the one previously obtained.
This is expected considering the arguments put forward in \citet{Owen2012b}, which is that the temperature at the sonic surface and hence the sound speed is fixed by the depth of the potential. At lower column densities a fixed temperature requires a lower ionization parameter for our new relations, thus yielding a higher mass loss rate as a consequence of the higher density.

\begin{table}
\caption{Mass-loss rates \label{tab:massLoss}}
\centering
\begin{tabular}{l c c}
\hline
\hline
\bf{simulation} & R$_\mathrm{hole}$ & $\mathbf{\dot{M}}$ \\
 & au & $10^{-8}$ M$_\odot$ yr$^{-1}$\\
\hline
PR & $0.33$ & $2.644$ \\
TR5 & $6.2$ & $2.940$ \\
TR10 & $9.9$ & $3.036$ \\
TR20 & $17.9$ & $2.923$ \\
TR30 & $28.2$ & $2.887$ \\
\hline
\end{tabular}
\end{table}

\begin{figure}
\centering
\includegraphics[width=.45\textwidth]{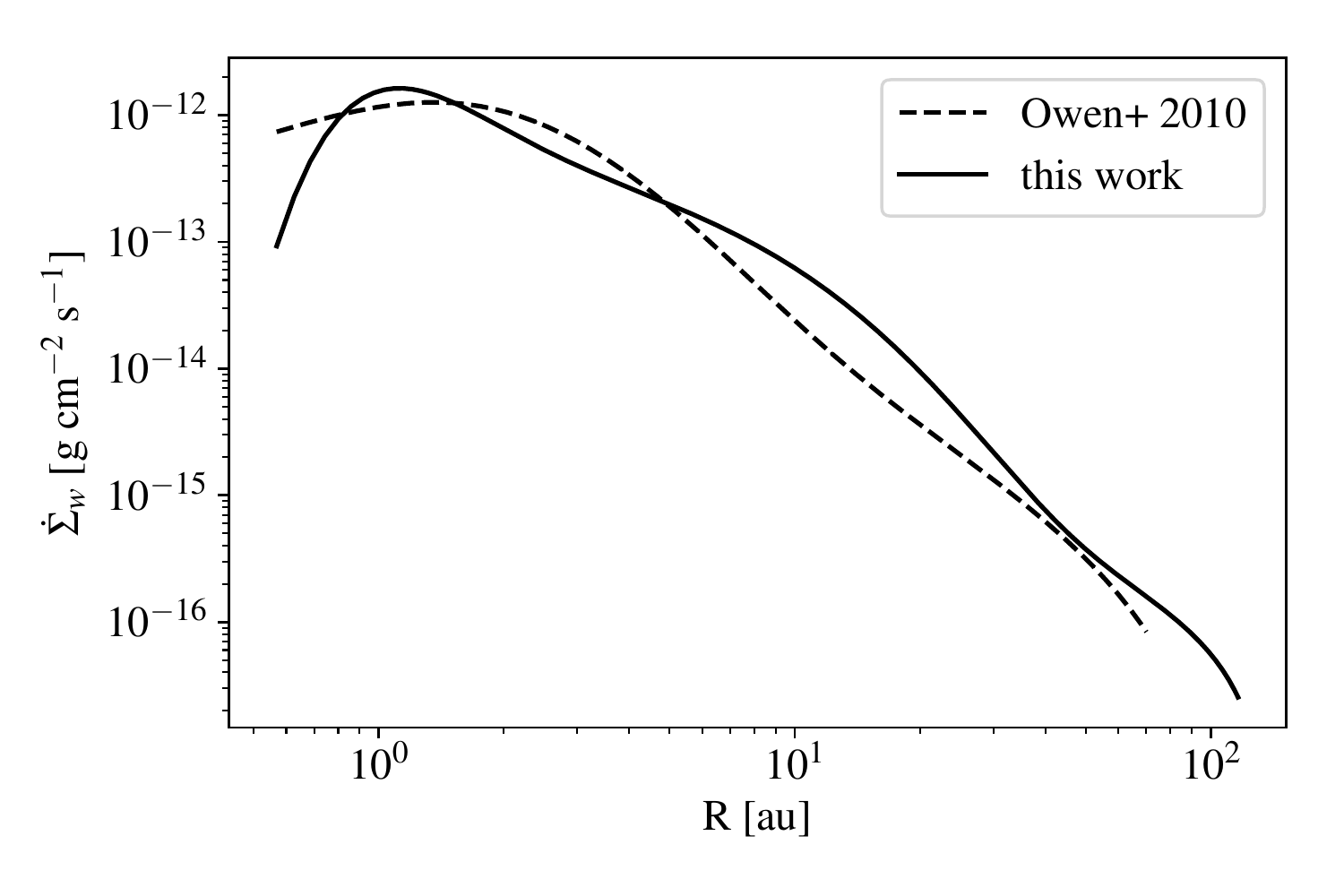}
\caption{Surface mass-loss profile for the primordial disc shown as the solid line, compared with the previous prescription by \citet{Owen2010} with the dashed line. \label{fig:Sigmadot1}}
\end{figure}

This difference results also in a modified surface density mass-loss profile (as shown in Fig.~\ref{fig:Sigmadot1}), where two peaks are visible around $1$ au and $20$ au.

The fit to the surface density mass-loss profile for the primordial disc is:

\begin{eqnarray}
\dot{\Sigma}_w(R) &= \ln{(10)} \bigg(\frac{6*a*\ln{(R)}^5}{R*\ln{(10)}^6} +
\frac{5*b*\ln{(R)}^4}{R*\ln{(10)}^5} +
\frac{4*c*\ln{(R)}^3}{R*\ln{(10)}^4} + \\ \nonumber
&\frac{3*d*\ln{(R)}^2}{R*\ln{(10)}^3} +
\frac{2*e*\ln{(R)}}{R*\ln{(10)}^2} + \\ \nonumber
&\frac{f}{R*\ln{(10)}}\bigg)
\frac{\dot{M}_w(R)}{2\pi R} \ [M_\odot\, {au}^{-2}\, {yr}^{-1}]\,
\end{eqnarray}
with
\begin{equation}
\dot{M}_w(R) = \dot{M}(L_X) 10^{a\log{R}^6 + b\log{R}^5 + c\log{R}^4 + d\log{R}^3 + e\log{R}^2 + f\log{R} + g}\,
\end{equation}
and a = -0.5885, b = 4.3130, c: -12.1214, d = 16.3587, e = -11.4721, f = 5.7248, g = -2.8562.
%a = -0.9478, b =  7.5586, c = -24.2529, d =  39.8226, e = -35.4372, f = 17.2032, g = -4.4011.
In the previous equations we use the standard convention and we refer to the natural logarithm with $\ln{}$ and the logarithm in base $10$ with $\log{}$. The radius is expressed in au, and $\dot{M}(L_X)$ is the normalization factor given by eq.~\ref{eq:MdotLx}.

\begin{figure*}
\centering
\includegraphics[width=.95\textwidth]{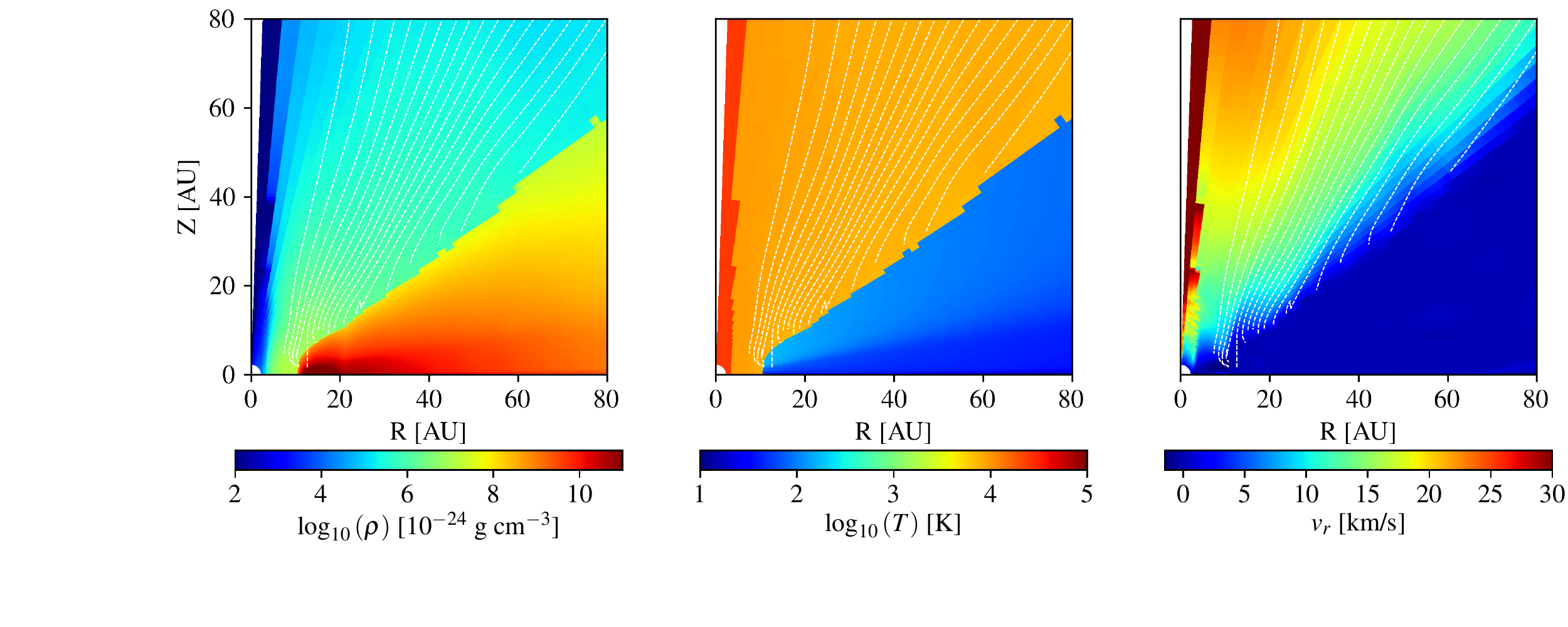}
\caption{Meridional plots for the TR15 simulation zoomed in close to the gap edge. Left panel: mass density; middle panel: temperature; right panel: radial velocity.
The streamlines are plotted with a dashed white lines at 5 per cent intervals of the integrated mass-loss rate. \label{fig:SigmaVelTempTrans}}
\end{figure*}

\subsection{Hole Radius dependence}

\begin{figure}
\centering
\includegraphics[width=.45\textwidth]{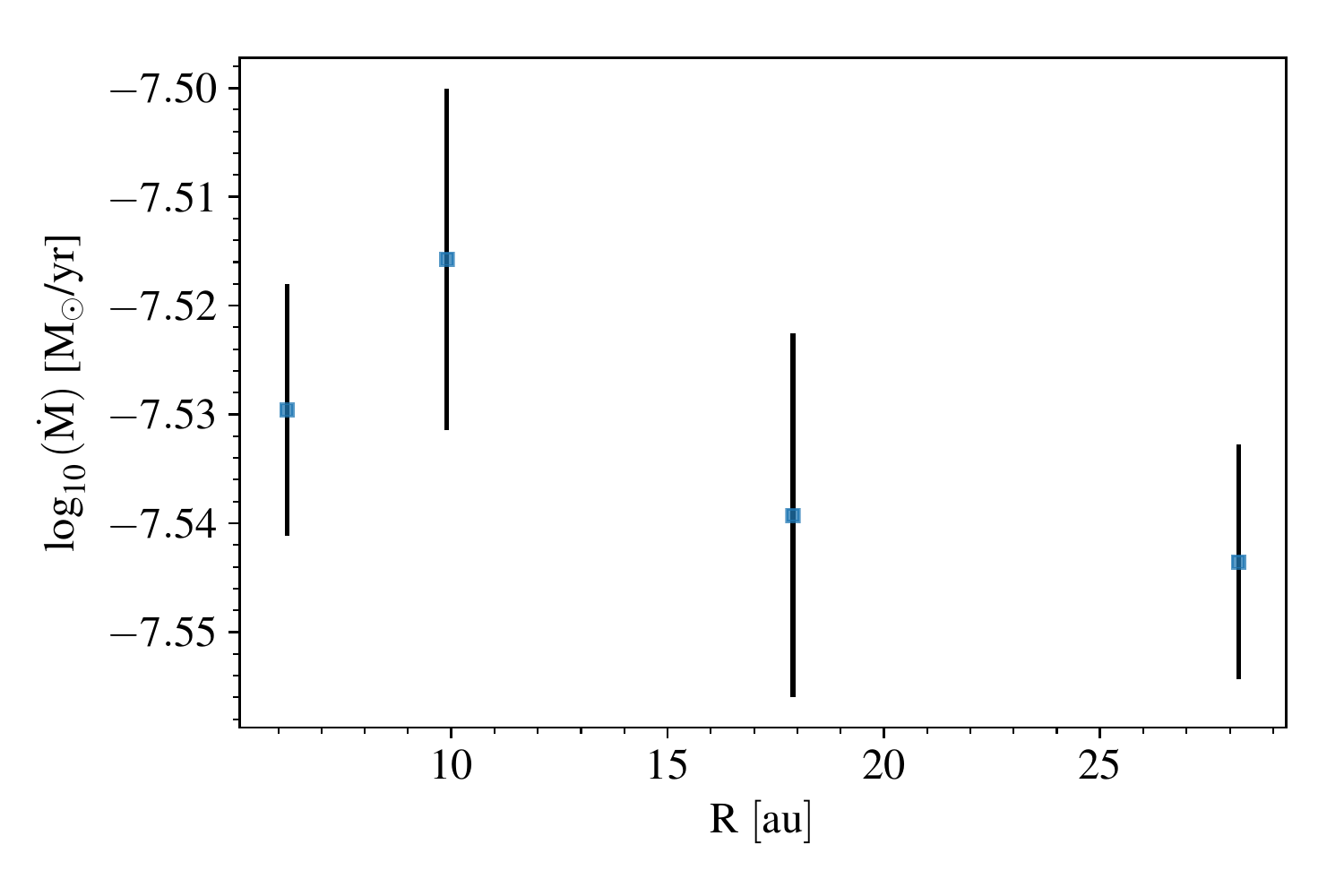}
\caption{Integrated mass-loss rate as a function of the inner hole for the transition discs. \label{fig:MdotLx}}
\end{figure}

As described in Section~\ref{sec:hydro}, we created a set of transition discs from the primordial disc and for the reference X-ray luminosity of \SI{2d30}{erg\ s^{-1}}, in order to test the dependence of the mass-loss rate due to photoevaporation as a function of the inner hole radii.
The result of our parameter study is reported in Tab.~\ref{tab:massLoss}, as well as in Figure~\ref{fig:MdotLx}.
The total mass-loss rate is increased with respect to the primordial disc, by a factor $\sim 1.12$, but it results independent of the inner radii of the disc. The surface density mass-loss profile for the transition discs are given by

\begin{equation}
\dot{\Sigma}_{w,T}(R) = a b^{x} x^{c-1} [x \ln(b)+c]\ \frac{1.12\, \dot{M}(L_X)}{2\pi R} \ [M_\odot\, {au}^{-2}\, {yr}^{-1}]\,
\end{equation}

where $x=(R-R_\mathrm{gap})$, a = \SI{1.1843d-1}, b = \SI{9.9695d-1}, c = \SI{4.8835d-1}, and $\dot{M}(L_X)$ is given by eq.~\ref{eq:MdotLx}.

%\begin{equation}
%\label{eq:MdotRhole}
%\log_{10}(\dot{M}(R_\mathrm{hole})) = A_\mathrm{M} -B_\mathrm{M}\cdot R_\mathrm{hole}
%\end{equation}

%where $A_\mathrm{M} = -7.309\pm0.009$, $B_\mathrm{M} = (9.592\pm0.573)\cdot10^{-3}$ where the error in the fit is given by the $1 \sigma$ interval reported in Fig~\ref{fig:MdotLx}, and it is coming from the small oscillations observed in the mass-loss rate evolution as discussed in Appendix~\ref{app:outerradius}.

\subsection{Stellar X-ray luminosity dependence}
The dependence of the integrated mass loss rate on the stellar X-ray luminosity was also investigated.

\begin{figure}
\centering
\includegraphics[width=.45\textwidth]{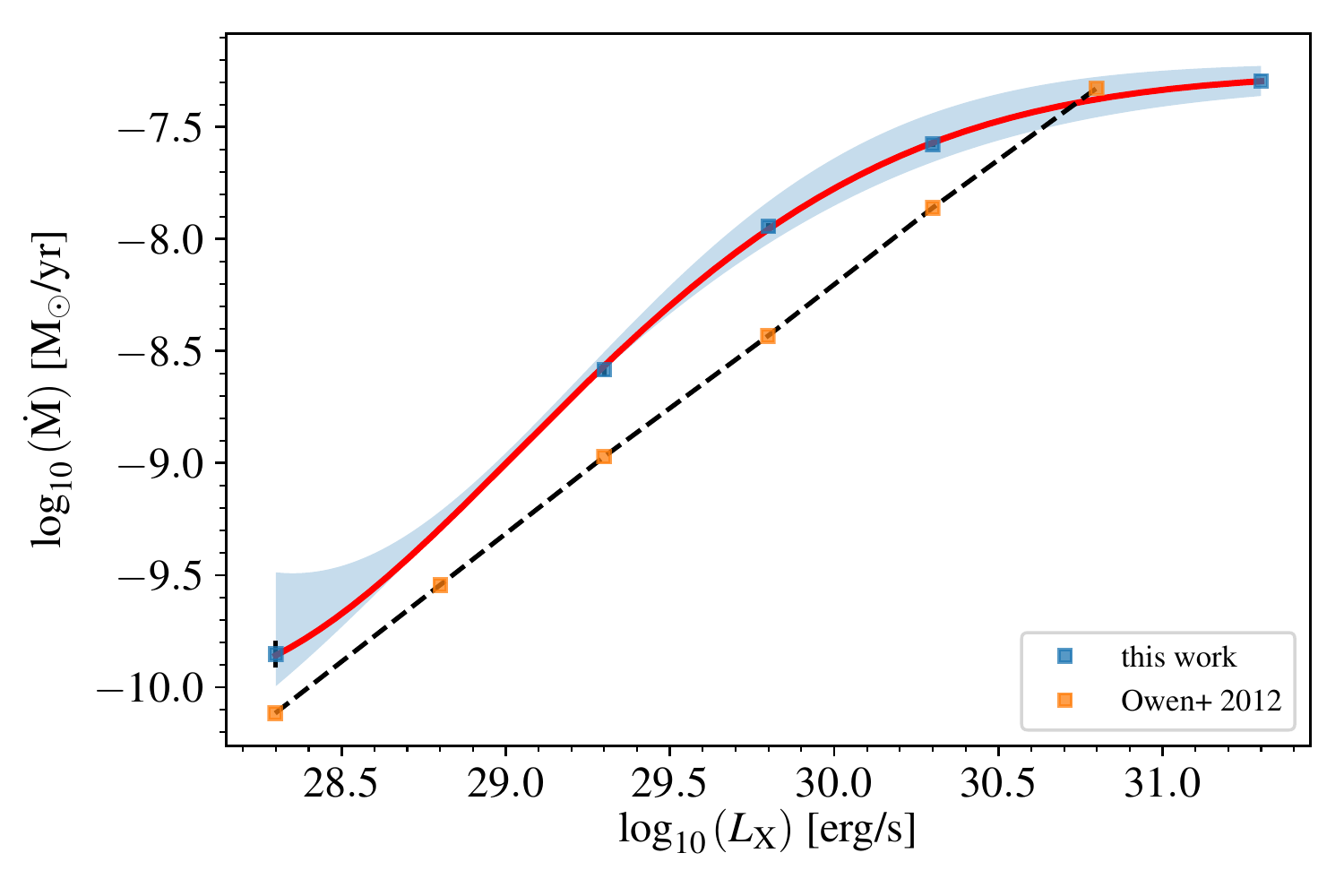}
\caption{Integrated mass-loss rate as a function of the X-ray star luminosity, from which the fit in eq.~\ref{eq:MdotLx} was derived. The relation from \citet{Owen2012b} is overplotted for comparison. \label{fig:MdotLx1}}
\end{figure}

\begin{figure}
\centering
\includegraphics[width=.45\textwidth]{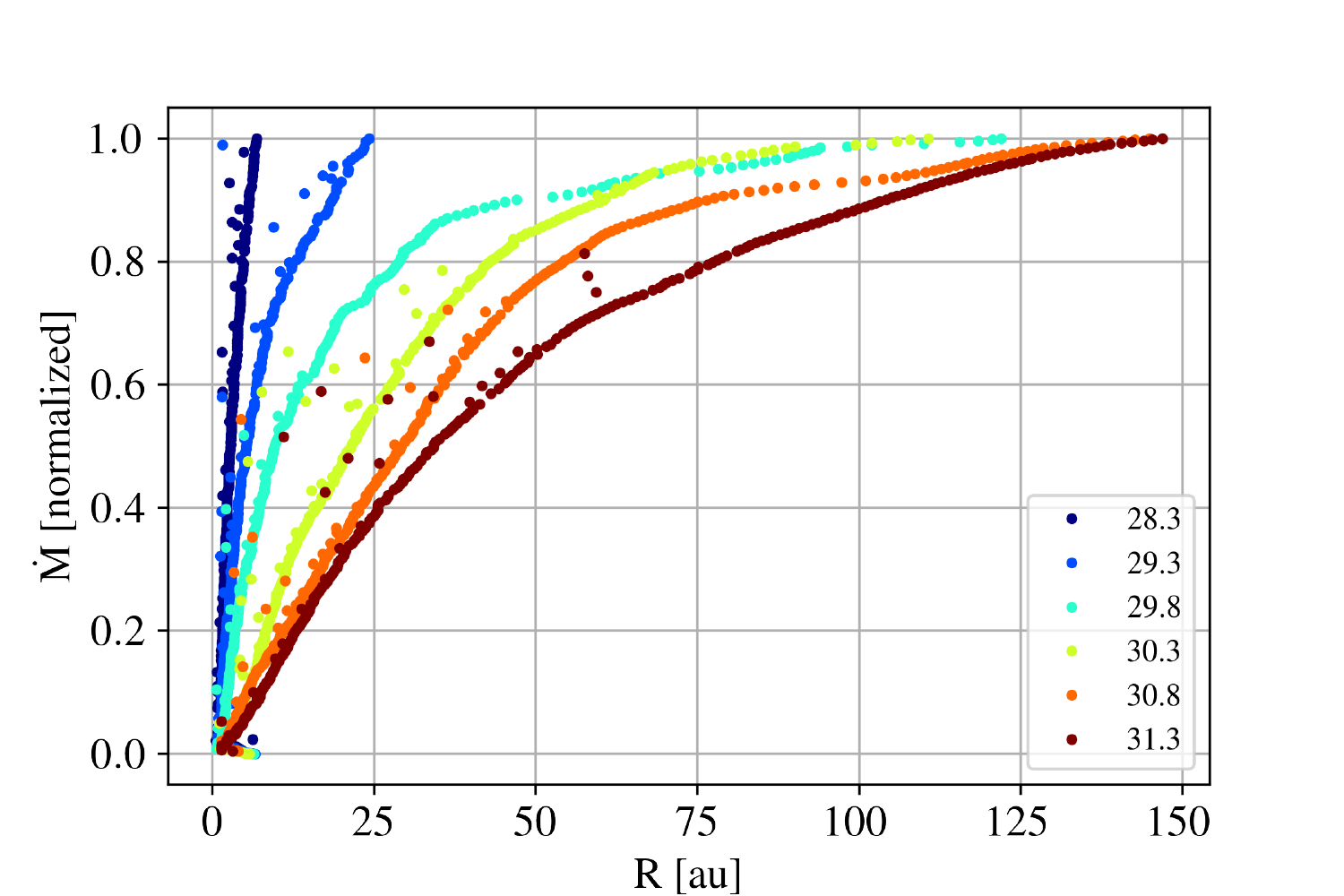}
\caption{Normalized cumulative mass-loss rate as a function of the distance from the host star for different X-ray luminosities (marked in log scale in the legend). \label{fig:LxComp}}
\end{figure}

The result is shown in Fig.~\ref{fig:MdotLx1} and it can be described by a sigmoidal function:

\begin{equation}
\label{eq:MdotLx}
%a / (1.0 + np.exp(-(x-b)/c))**d
%\log_{10}(\dot{M}(L_X)) = \frac{A_\textrm{L}}{(1 + \exp{[-(\log_{10}(L_X)-B_\textrm{L})/C_\textrm{L}})])^{D_\textrm{L}}} \,
\log_{10}(\dot{M}(L_X)) = A_\textrm{L} * \exp{\left[\frac{(\ln{(\log_{10}(L_X))}-B_\textrm{L})^2}{C_\textrm{L}}\right]} + D_\textrm{L},
\end{equation}

in $M_\odot\, \mathrm{yr}^{-1}$, with $A_\textrm{L} = -2.7326$, $B_\textrm{L} = 3.3307$, $C_\textrm{L} = -2.9868\cdot10^{-3}$, $D_\textrm{L} = -7.2580$.
%$A_\textrm{L} = -7.080\pm0.001$, $B_\textrm{L} = 30.488\pm0.001$, $C_\textrm{L} = 0.273\pm0.001$, $D_\textrm{L} = -0.0395\pm0.0003$.
An interesting behaviour is observed, where for high luminosities, the integrated mass-loss rate tends to saturate around \SI{1d-7}{M_{\odot} yr^{-1}}.
%This trend can be linked to the relation between the ionization parameter and the gas temperature (from Figure~\ref{fig:tempxi}), where for high ionization parameters the maximum temperature does not increase considerably.
%Thus, stars with high X-ray luminosities will ionize a greater fraction of the circumstellar disc, but this will not result in a stronger wind.
At high X-ray luminosities, only the flat region of the $T-\xi$ relation (see Fig.~\ref{fig:tempxi}) is accessible to the flow, and the theory from \cite{Owen2012a} breaks down. In this regime, the disc jumps straight from the disc temperature to \SI{1d4}{K}, and the depth of the potential is irrelevant, but the mass-loss rate is set solely by the disc structure.
Thus the surface-mass loss profile becomes dependent on the X-ray luminosity as shown in Fig.~\ref{fig:LxComp}, in contrast with Fig. 4 of \citet{Owen2011}. Moreover, the disc structure is affected by the higher X-ray luminosity and the area accessible by the X-ray radiation is increased with higher X-ray luminosities since the disc is more puffed up.

\subsection{Transition Disc demographics}

Photoevaporation models, naturally produce transition discs as a result of inside-out dispersal. Other processes, including giant planet formation, could also produce dust-depleted cavities in discs. It is thus important to differentiate amongst the possible origins of observed transition discs, if the latter are to be used as laboratories to study disc dispersal and/or planet formation. The observed transition disc demographics span a large parameter range in terms of the size of the central cavity (from sub-au to hundreds of au) and the measured accretion rate onto the central star (from a few non-accreting sources to accretion rates similar to those of full discs). This points to the notion of transition discs being a very diverse class of objects (e.g. \citep[e.g.][]{Owen2012a,Owen2016,Ercolano2017}, where the X-ray photoevaporation models of \citet{Owen2010,Owen2011}, can only explain a fraction of the observed objects, namely discs with small (a few \si{\au}) cavities and low accretion rates (less than \SI[mode=text]{1d-8}{\Msun.\yr^{-1}}).

\begin{figure}
\includegraphics[width=0.45\textwidth]{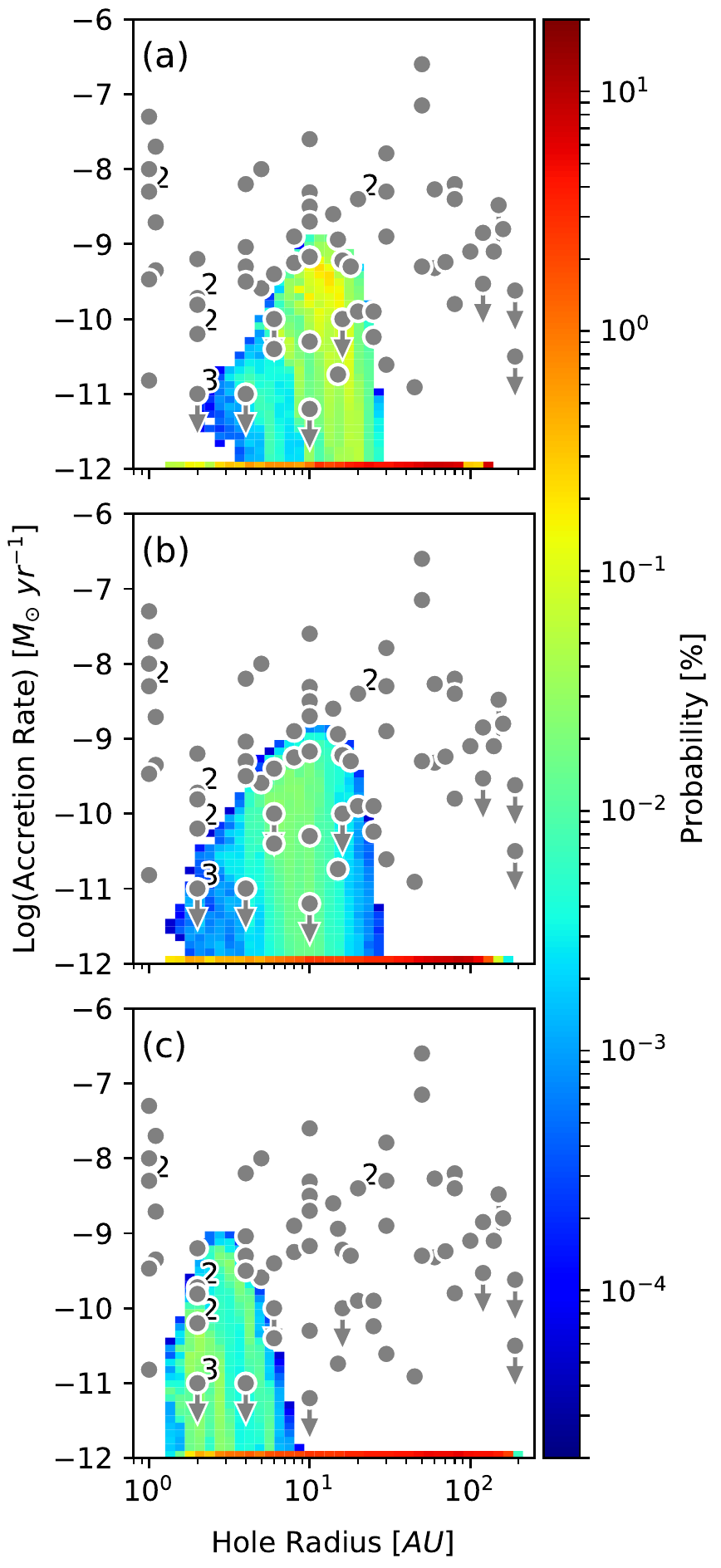}
\caption{Transition disc demographics of our synthetic populations. The grey circles show observed transition discs. Numbers next to the circle indicate the total number of sources at that point, arrows indicate that the accretion rate is an upper limit. The colored areas show the probability to find a transition disc with the corresponding accretion rate and hole radius, calculated from populations of our X-Ray driven photoevaporating disc models. Discs with an accretion rate lower than \SI{1d-12}{M_{\odot} yr^{-1}} are shown at the bottom. Panel a shows the results for our disc population with fixed initial parameters, panel b shows the results for the population with variable scaling radius $R_1$ and panel c is a reproduction of the population in \citet{Owen2011}. \label{fig:popsynts}}
\end{figure}

The wind profile has a strong influence on the evolution of the surface density of the disc \citep{Ercolano2015,Ercolano2018,Jennings2018}, for that reason, we have explored here the consequences of our improved X-ray photoevaporation model on the transition disc demographics. In Figure \ref{fig:popsynts} we show the results of our transition disc population syntheses (panel a) and compare it to the calculations of \citet{Owen2011}, shown in panel c of the same figure. The observations are represented by grey points and exclude stars of spectral type A, G and earlier \citep[see][for the individual references to the observational data]{Ercolano2017}.

We refer the reader to \citet{Owen2011} for details of the methods employed, which is based on one-dimensional viscous evolution models of discs under the effect of photoevaporation \citep[see also][for details of the code used for the calculations]{Ercolano2015}. We sample the X-ray luminosity function of the central stars stochastically using observational data for the Taurus region. We use a cumulative X-Ray luminosity function built from observed luminosities in the range \SIrange[range-units = brackets]{0.3}{10}{\keV} of pre-main sequence stars with masses $0.5 \leq M \leq \SI{1.0}{M_{\sun}}$ in the Taurus cluster \citep{Gudel2007}. In order to compare with \citet{Owen2011} we also initially assume a single disc set up consisting of a viscosity parameter, $\alpha = \num{2.5d-3}$ and initial disc scaling radius, $R_1 = \SI{18}{\au}$.

The new calculations show a better overlap with the observations. Accreting transition discs with cavities up to \SI{\sim 30}{\au} are predicted by the new models, compared to \SI{< 10}{\au} for the \citet{Owen2010,Owen2012b} models. The reason for that is that the new profiles are more efficient at removing material at larger disc radii, resulting in the more efficient outward dispersal of the outer disc, whose inner edge can then reach larger radii before the inner disc is viscously accreted onto the star and the accretion signature is lost.

Our new models however still fall short of being able to explain the several observed objects with larger inner cavities and more vigorous accretion rates. The problem persists also if we relax the assumption of all sources having the same $\alpha$ and $R_1$ and follow instead the approach of \citet{Ercolano2018} by randomly sampling the initial scaling radius, $R_1$ from a uniform distribution allowing $R_1$ to vary between \SIrange[range-units = single]{10}{100}{\au}. In these models, $\alpha$ is then chosen accordingly in order to ensure that the dispersal timescale of the populations match observations \citep{Gudel2007}. Assuming a simple inverse dependence of $\alpha$ on $R_1$ one finds:
\begin{equation}
\alpha = 2 \cdot 10^{-3} + 7.5 \cdot 10^{-5} \cdot R_1.
\end{equation}
The resulting population synthesis is shown in panel b of Figure \ref{fig:popsynts}, where we see that the relaxation of initial disc conditions leads to an increase in the inner cavity sizes of accreting transition discs that can be covered by the models up to \SI{30}{au}. However, as mentioned above, these models can still only explain half of the observed sources.

Large cavity, strongly accreting transition discs are often taken as possible signposts for (multiple) giant planet formation \citep[e.g][]{Zhu2011,Dodson2011} although this interpretation is not supported by theoretical models for many individual sources \citep[e.g.][]{Zhu2012,Owen2014,Owen2016} and also on statistical grounds \citep[e.g.][]{Dong2016}.
\citet{Ercolano2018} provide an alternative explanation, where large cavity, strongly accreting transition discs are a result of X-ray photoevaporation in metal (C and O) depleted discs. This interpretation, however, needs to be confirmed by more rigorous hydrodynamic modelling at variable metallicities, which is the focus of a forthcoming paper (W\"olfer et al. 2019, in prep.).
An alternative explanation for the large accretion rates measured in large cavity TDs was also presented by \citet{Wang2017b}, involving a magnetically driven supersonic accretion of low surface density material. \citet{Ercolano2018} however argue that this  model is at odds with recent observations from the Atacama Large Millimiter Array (ALMA) of the inner cavities of some of these objects, showing that a large quantity of gas is present inside the dust cavity \citep{Bruderer2014,vanderMarel2015,vanderMarel2016,Dong2017}.

Finally, our models predict a large population of non-accreting transition discs with large cavities, which were named 'relics' by \citet{Owen2011} and have not yet been observed. \citet{Owen2011} predict that $\sim 97 \%$ of all transition discs should be relics. Our new models predict $\sim 88.8 \%$ and $\sim 96.4 \%$ of relics when fixed and variable initial disc parameters are assumed, respectively. As final thermal sweeping seems to be ruled out by recent calculations \citep{Haworth2016}, this discrepancy likely points to additional processes taking over disc dispersal at large radii (e.g. FUV-photoevaporation or enhanced X-ray photoevaporation at large radii in C-depleted sources, W{\"o}lfer et al., in prep., Owen \& Kollemeier, in prep.).

\subsection{The thermal and ionisation state of the wind and its observables}

\citet{Ercolano2010,Ercolano2016} presented a study of possible wind tracers, including the [Ne~II] fine structure line at \SI{12.8}{\micro m} and a number of collisionally excited lines (CELs) from atomic and low ion species like C, O and S. These lines were targeted in those studies as they are often detected with few km/s blue-shifts in observational surveys \citep[e.g.][]{Rigliaco2013,Natta2014, Simon2016}, thus bearing the signature of a slow-moving disc wind which may be photoevaporative in nature. While \citet{Ercolano2010,Ercolano2016} show that X-ray photoevaporation models can indeed reproduce many of the reported observations \citep[but see also][]{Simon2016}, \citet{Ercolano2016} also demonstrate that CELs, with their strong temperature dependence, are not suitable as measures of mass-loss rates as they just trace a small heated volume of gas high in the wind rather than its bulk at its base \citep[see also discussion in][]{Ercolano2017}. Nonetheless, the ability to reproduce the observed line intensities and their approximate profiles is an important requirement for a viable photoevaporation model.

\begin{figure*}
\includegraphics[width=\textwidth]{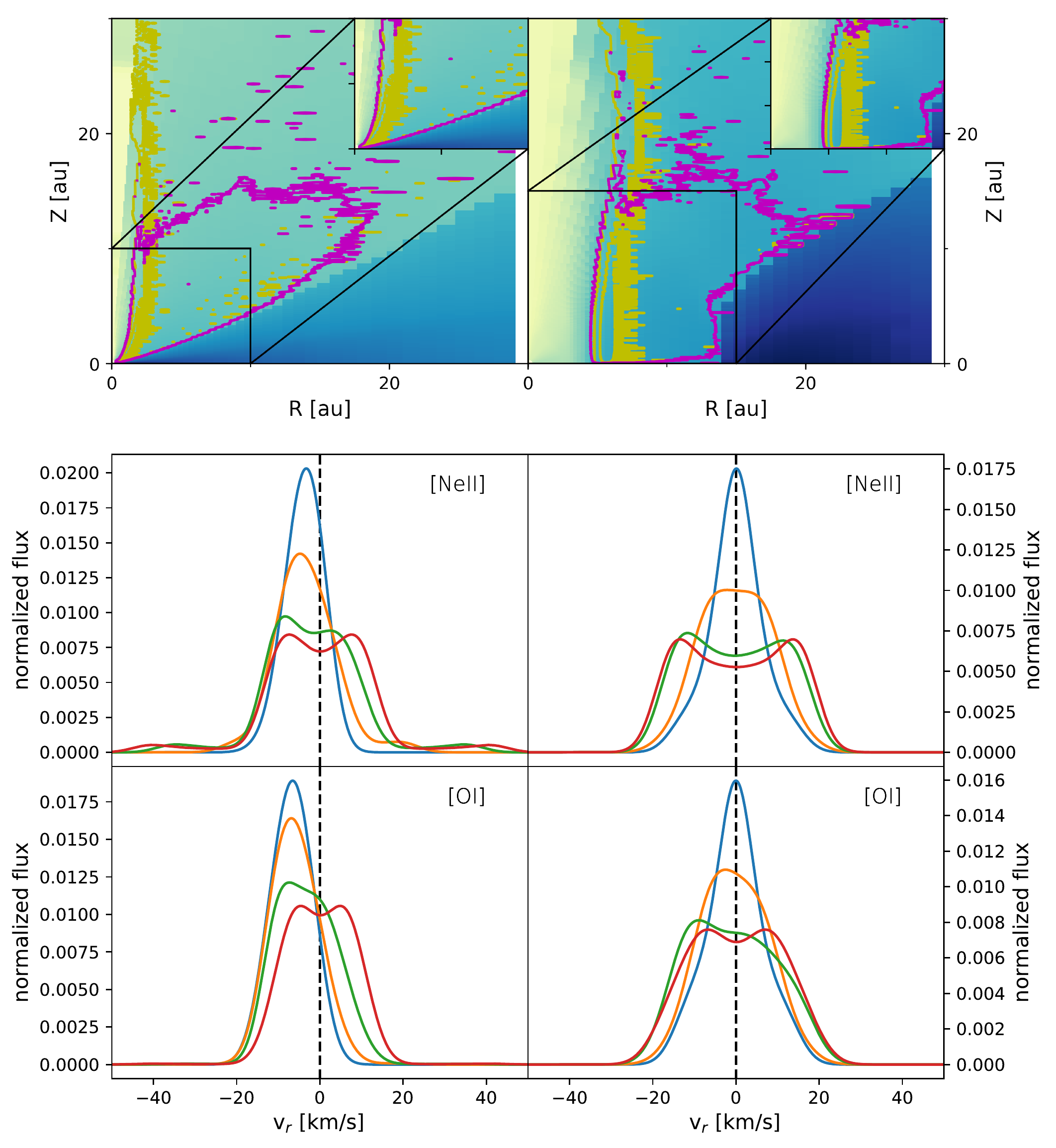}
\caption{The top panels show density maps for the PR and TR15 simulations, where superimposed is the location of the $85\%$ emission region of the [OI] \SI{6300}{\angstrom} line (yellow contour) and [NeII] \SI{12.8}{\micro m} line (purple contour). The lower panels show the emission line profiles for the same lines at four different disc inclinations: $0^{\circ}$ (blue line), $30^{\circ}$ (yellow line), $60^{\circ}$ (green line) and $90^{\circ}$ (red line). \label{fig:lines}}
\end{figure*}

One example is the observation of a blue-shift in the [OI] line at \SI{6300}{\angstrom}, which immediately points to a warm quasi-neutral wind, hence ruling out EUV-driven photoevaporation, which produces a fully ionised wind, as in the case of the recent simulations by \citet{Wang2017}.
In this context, we have calculated here line intensities and line profiles from our new models, via a post-processing of the hydrodynamical grids through the 3D photoionisation and dust radiative transfer code {\sc MOCASSIN} \citep{MOCASSIN1,MOCASSIN2,MOCASSIN3}, and showed that these are in agreement with current observational data.
The results are summarised in Fig.~\ref{fig:lines} and in Tab.~\ref{tab:lines}.
On the upper-left panel the surface density of the primordial disc (PR) is shown, with superimposed the emission regions of [OI] \SI{6300}{\angstrom} in yellow, and [NeII] \SI{12.8}{\micro m} in purple.
In the middle and low-left panels the relative emission lines are then displayed for $4$ different inclinations ($0^{\circ}$, $30^{\circ}$, $60^{\circ}$, and $90^{\circ}$).
A zoom-in onto the innermost disc is presented in upper plots, in order to show that the emission regions becomes very close to the bound material in the disc, reducing the blue-shift of the produced lines.
As a result, the peak for the [OI] in the edge-on configuration is significantly blue-shifted while, for the [NeII], the more extended bound component is reducing the blue-shift to only a few km/s (see also Tab.~\ref{tab:lines}).
On the right-hand side of Fig.~\ref{fig:lines} the same plots are shown for a transition disc (TR10), where the profiles are much more symmetric since both sides of the disc are visible through the central hole, thus allowing the detection of the red-shifted as well as the blue-shifted portion of the wind.
A more detailed study of spectroscopic wind diagnostics predicted by our new models, compared with observations, will be presented in an forthcoming paper (Ercolano \& Picogna, in prep.).

\begin{table}
\centering
\begin{tabular}{lcccc}
\hline
& Ne~II \SI{12.8}{\mu m} & & [OI] \SI{6300}{\angstrom} & \\
\hline
i & v$_\mathrm{peak}$ & FWHM & v$_\mathrm{peak}$ & FWHM \\
$^{\circ}$ & km s$^{-1}$ & km s$^{-1}$ & km s$^{-1}$ & km s$^{-1}$\\
\hline
\hline
$0$ & $-3.56 \pm 0.01$ & $11.50$ & $-6.707 \pm 0.006$ & $12.427$ \\
$30$ & $-3.94 \pm 0.03$ & $15.94$ & $-6.28 \pm 0.02$ & $14.22$ \\
$60$ & $-1.93 \pm 0.06$ & $18.15$ & $-3.21 \pm 0.05$ & $15.97$ \\
$90$ & $0.10 \pm 0.04$ & $8.25$ & $0.11 \pm 0.01$ & $16.36$ \\
\hline
\end{tabular}
\caption{Line profiles from primordial discs irradiated by $\log_{10}{(L_\mathrm{X})}$ = \SI{30.3}{erg\ s^{-1}}. The table lists the velocity of the peak and the full width at half-maximum (FWHM) for $4$ inclinations from $0$ to \SI{90}{^\circ}. The profiles were degraded to an instrumental resolution of $R=100,000$. \label{tab:lines}}

\end{table}

\section{Conclusions}\label{sec:conclusions}

In this work, we performed hydrodynamical models of an X-EUV-driven photoevaporative wind for a primordial disc and transition discs with different inner holes.
We improved on the models of \citet{Owen2010,Owen2011,Owen2012b} by parameterising the temperature as a function not only of local properties of the gas (via the ionization parameter) but also on the column density to the star.
This new approach resulted in new mass-loss profiles and total mass-loss rate of \SI{2.644d-8}{M_\odot yr^{-1}} (for a star X-ray luminosity of \SI{2d30}{erg s^{-1}}).
The main results from our models can be summarised as follows:
\begin{itemize}
\item the mass-loss rates generated by photoevaporative winds flattens for high stellar X-ray luminosities (Figure~\ref{fig:MdotLx1}), reflecting the flattening of the temperature profile for high ionisation parameters (see Figure~\ref{fig:tempxi});
\item the mass-loss rate does not depend on the inner-hole radii for transition discs (Figure~\ref{fig:MdotLx});
\item the demographics of transition discs obtained using the new scaling laws can explain a larger fraction of the observed transition discs (Figure~\ref{fig:popsynts}), and in general it can explain accreting transition discs with cavities up to \SI{30}{\au};
\item the line profiles generated from the hydrodynamical models can reproduce the two-components profile found by observational surveys (Figure~\ref{fig:lines}), where the peak at $\theta = 0$ is given by bound material in the inner part of the protoplanetary disc. These models, however, cannot reproduce the blueshift in the broad component that is often reported in the observations \citep{Simon2016,Banzatti2018}.
\item the interaction between the photoevaporative wind and the outer boundary of the computational domain needs to be taken with caution, since it can generate spurious numerical oscillation, affecting the final mass-loss rate calculation (as observed in \citet{Wang2017}).
\item as observed in Appendix~\ref{app:Owen}, the differences in the cumulative mass-loss rate and surface density profiles are limited when comparing our numerical set-up with the one of \citet{Owen2010}, pointing out that the results are quite robust. Nevertheless, the dependence of the temperature profile on the column density, leads to an increased mass-loss rate at large radii (see Fig.~\ref{fig:CumMdot}) and to the observed higher fraction of transition discs with large holes in the population synthesis models (see Fig.~\ref{fig:popsynts}).
\item even though we include EUV in the spectrum irradiating the planet-forming disc, this cannot dominate the total mass-loss rate (as proposed by \citet{Wang2017}) because, as observed also in \citet{Owen2010}, the EUV radiation field is absorbed for small column densities and it is not able to reach the high-density parts of the disc, thus its contibution to the total mass loss rate is negligible
\item \citet{Wang2017} found adiabatic cooling to be an important channel to the thermal balance in their simulations. We find, on the contrary (and in agreement with previous calculations by \citet{Owen2010} that this can be neglected, given that the energy flux gained in the wind is always much smaller ($< 8 \%$) that from the XEUV spectrum. Thus, we conclude that the gas is approximately in thermal equilibrium. Further to this point, we performed an explicit comparison of the advection and recombination timescales throughout our grids, finding the former to significantly exceed the latter (see App.~\ref{app:radeq}).
Rather than advection, we suspect that a more likely source of discrepancy between the temperature structure of the models of \citet{Wang2017} and ours \citep[and][]{Owen2010}, is the limited sample of frequency points ($6$) used in their models (we use circa $1000$), which is unable to describe the ionisation structure of the gas.

\end{itemize}
The models presented here form the basis for future investigations aiming at finally being able to perform quantitative spectroscopy of disc winds from observations. Work is already in progress on the next steps, which include a study of the metallicity dependance of on the mass-loss rates (W{\"o}lfer et al., in preparation), detailed calculation of synthetic spectra from our calculations (Ercolano \& Picogna, in preparation), and the inclusion of a realistic chemical network to determine the importance of FUV heating compared to the XEUV component in the different regions of the disc.

\section*{Acknowledgements}
We thank the anonymous referees for the useful comments and suggestions which substantially helped improving the quality of the paper.
GP acknowledges support from the DFG Research Unit ``Transition Disks'' (FOR~2634/1, ER~685/8-1).
JEO is supported by a Royal Society University Research Fellowship. \\
We thank T. Haworth for the thermal sweeping models, T. Grassi for the insightful discussions, and L. W{\"o}lfer for the help with the MOCASSIN fits.\\
This work was performed partly on the computational resource ForHLR I funded by the Ministry of Science, Research and the Arts of Baden-W\"{u}rttemberg and the DFG, and partly on the computing facilities of the Computational Center for Particle and Astrophysics (C2PAP).

%%%%%%%%%%%%%%%%%%%%%%%%%%%%%%%%%%%%%%%%%%%%%%%%%%

%%%%%%%%%%%%%%%%%%%% REFERENCES %%%%%%%%%%%%%%%%%%

\bibliographystyle{mnras}
\bibliography{references}

%%%%%%%%%%%%%%%%%%%%%%%%%%%%%%%%%%%%%%%%%%%%%%%%%%

%%%%%%%%%%%%%%%%%%%% APPENDICES %%%%%%%%%%%%%%%%%%

\appendix

\section{Comparison test with previous model}\label{app:Owen}
\begin{figure}
\includegraphics[width=\columnwidth]{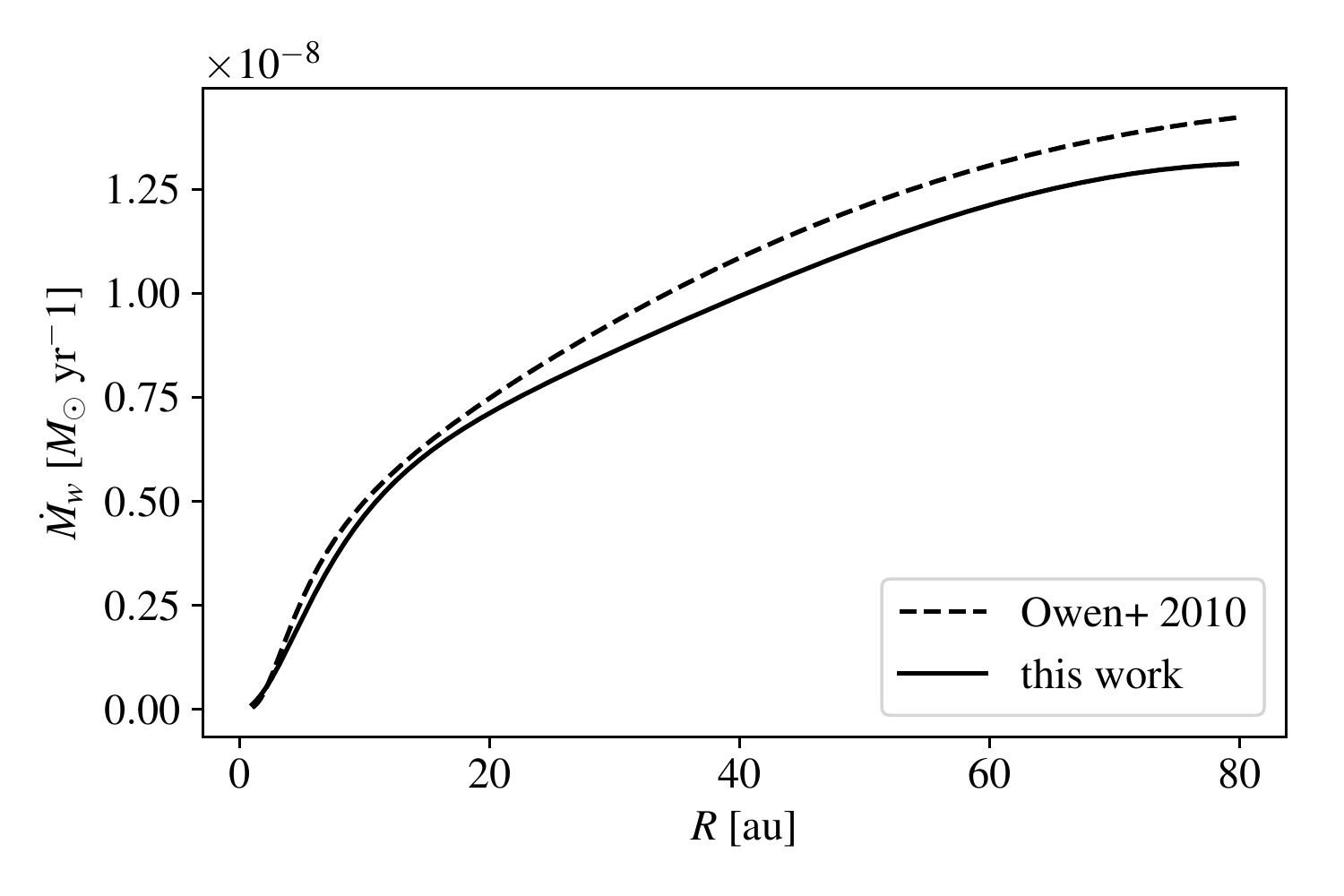}
\includegraphics[width=\columnwidth]{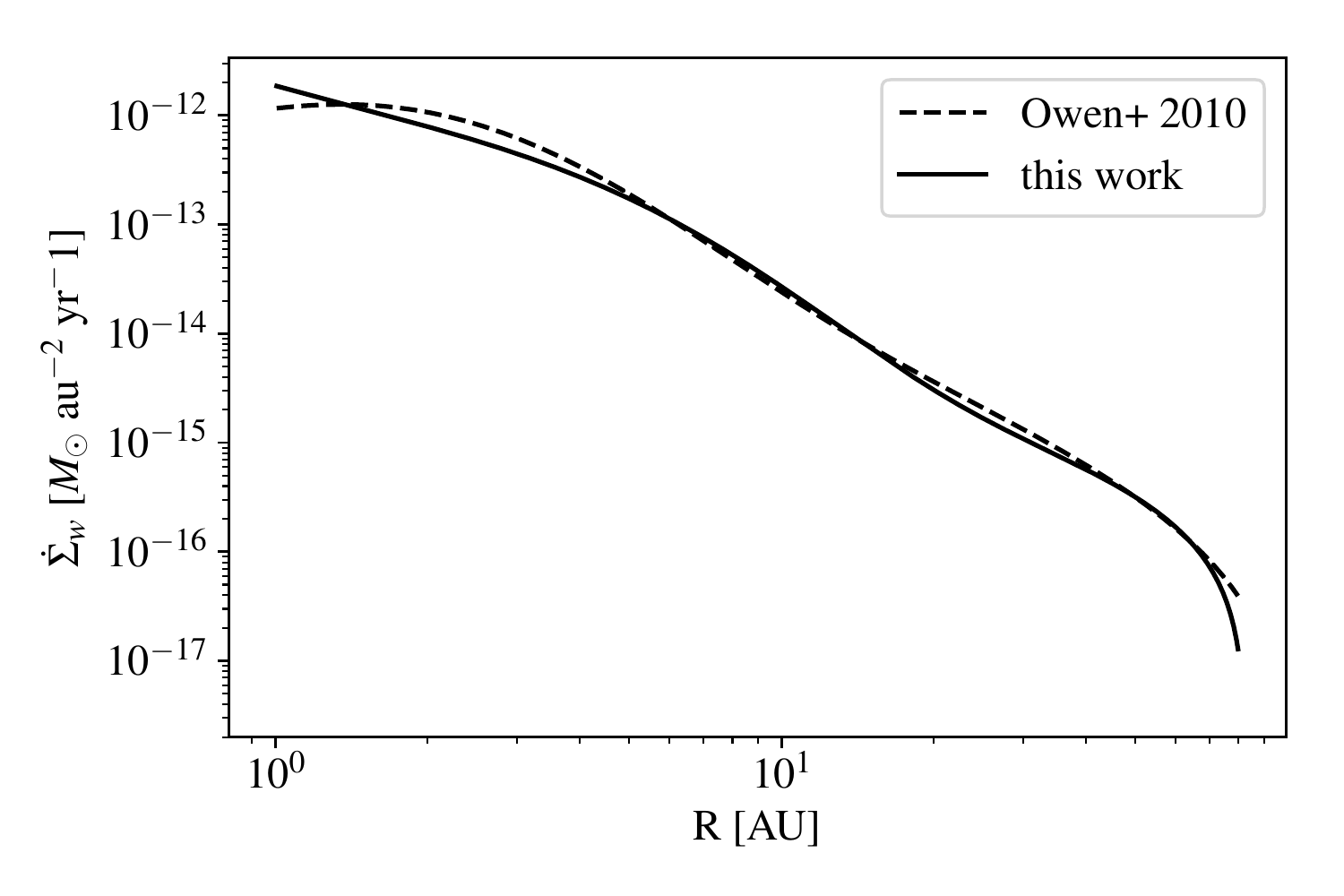}
\caption{Comparison of the cumulative mass-loss rate (top panel) and surface mass-loss profile (bottom panel) obtained with the method of \citet{Owen2010} with their original results. \label{fig:testOwen}}
\end{figure}

This study is an improvement over the series of models made by \citet{Owen2010}. As a first step we recreate the set-up of \citet{Owen2010} for their standard case (a primordial disc with $\log_{10}{(L_x)} = 30.3$) and a column-density independent temperature prescription.
In Fig.~\ref{fig:testOwen} we show the direct comparison of the density, temperature and radial velocity structure, where no major difference is observed both in the cumulative value of the mass-loss rate, and in the surface density mass-loss rate.

\section{Radiative equilibrium}\label{app:radeq}
\begin{figure}
\includegraphics[width=\columnwidth]{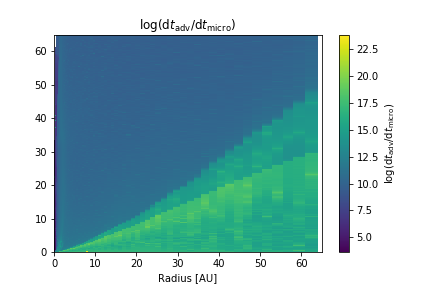}
\caption{\label{fig:testdt}}
\end{figure}

The approach used in this paper is based on the assumption of radiative equilibrium within the disc. This condition is met when the micro-physical time-scale, affecting the temperature equilibrium, is shorter than the hydrodynamical time-scale.
Usually, hydrogen recombination is the microphysical process with the longest time-scale \citep{Ferland1979,Salz2015}
\begin{equation}
  \tau_\mathrm{rec} = \frac{1}{\alpha_\mathrm{A}(T_\mathrm{e})n_\mathrm{e}} \simeq 1.5\cdot10^9 T_\mathrm{e}^{0.8}n_\mathrm{e}^{-1} \,,
\end{equation}
where $T_\mathrm{e}$ is the electron temperature, $n_\mathrm{e}$ is the electron density, and $\alpha_\mathrm{A}(T_\mathrm{e})$ is the temperature-dependant recombination rate.
Thus, we did an a-posteriori check of the whole computational domain in order to check whether the hydrodynamical time-scale was longer than the hydrogen recombination time-scale in the regions important for the wind dynamics
\begin{equation}
  \tau_\mathrm{adv} > \tau_\mathrm{rec}.
\end{equation}
The result is shown in Fig.~\ref{fig:testdt}, where the fraction of the advection and recombination timescales is plotted.
In the entire computational domain the hydrodynamical timescale is several order of magnitudes greater than the microphysical one, showing that our assumption is valid.

\section{Choice of the disc outer radius}\label{app:outerradius}
\begin{figure}
\includegraphics[width=\columnwidth]{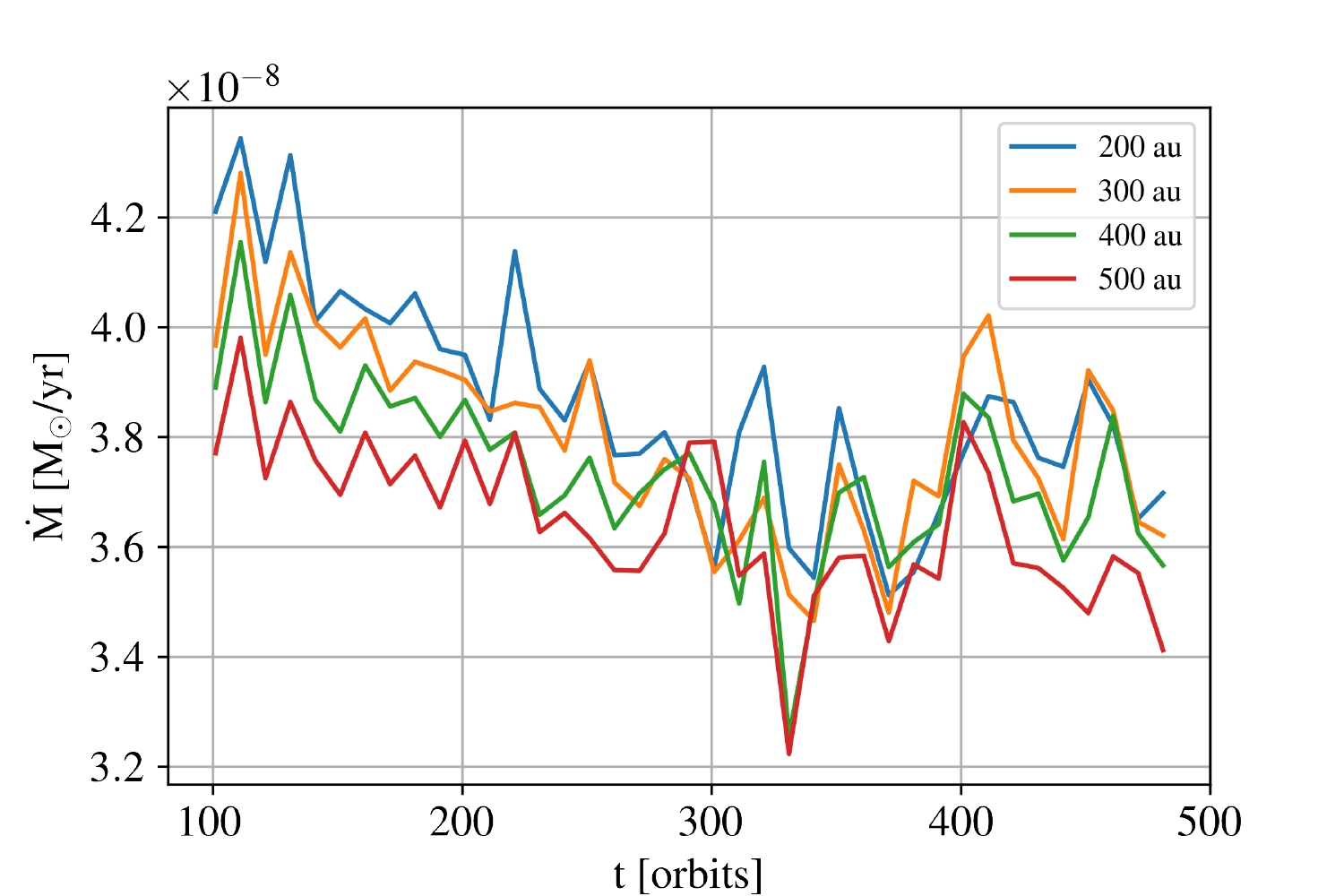}
\caption{Total mass-loss rate evolution as a function of time calculated from different outer radius up to the disc surface.\label{fig:radii}}
\end{figure}

The choice of the disc outer radius is important to discriminate between the material that is removed from the disc and the one that is simply redistributed within it.
In principle, we can use our hydro model self-consistently to determine if the material becomes unbound or not.
However, in order to have stable streamlines at hundreds of au from the central star, we need many more orbits than the ones modelled.
We present in Fig.~\ref{fig:radii} the evolution of the total mass loss rate for the primordial disc considering the outer radius at the different locations. We see that in the first hundred orbits the mass-loss rate is increased for lower outer radii since we are considering streamlines that are falling back towards the disc mid-plane at further locations. Nevertheless, as the disc reaches a quasi-stable equilibrium, the differences become smaller.
We decided to use the smallest radii which was not cutting out regions where the photoevaporation was effective, in order to have streamlines as stable as possible, thus we choose 200 au as our outer radius.

\bsp
\label{lastpage}
\end{document}